# Benchmarking Resource Usage of Underlying Datatypes of Apache Spark

Brittany Nicholls, Mariama Adangwa, Rachel Estes, Hugues Nelson Iradukunda, Qingquan Zhang, Ting Zhu
Department of Computer Science and Electrical Engineering
University of Maryland, Baltimore County
Email: {brn1, bama4, jd28691, hiraduk1, q, zt}@umbc.edu

*Abstract*—The purpose of this paper is to examine how resource usage of an analytic is affected by the different underlying datatypes of Spark analytics - Resilient Distributed Datasets (RDDs), Datasets, and DataFrames. The resource usage of an analytic is explored as a viable, and preferred alternative of benchmarking big data analytics instead of the current common benchmarking performed using execution time. The run time of an analytic is shown to not be guaranteed to be a reproducible metric since many external factors to the job can affect the execution time. Instead, metrics readily available through Spark including peak execution memory are used to benchmark the resource usage of these different datatypes in common applications of Spark analytics, such as counting, caching, repartitioning, and KMeans.

*Index Terms*—apache,spark,benchmark,analytics,big data

## I. INTRODUCTION

Apache Spark is a common analytics engine used in processing big data. It claims that a Spark job can run 100% faster than a similar Hadoop MapReduce job. [9] While this is an impressive metric, it is a bit misleading because it fails to capture the fact that Spark jobs mostly operate in main memory whereas Hadoop MapReduce jobs do not. It is common knowledge that accessing data in main memory is much faster than trying to move data from a hard drive to main memory, so it seems natural that a tool that utilizes main memory much more heavily would be faster than a tool that does not. This leads to questions such as: How much memory does a Spark analytic need to do the same thing as a Hadoop MapReduce analytic? Is purchasing servers with a lot of memory worth the benefits of a lower execution time? These types of questions are becoming more important since common cluster tools, such as AWS, are charging people for the resources they use, including memory.

Thus, in this paper we are going to analyze the resource usage of different Spark jobs. The primary focus will be on how the underlying datatypes - Resilient Distributed Datasets (RDDs), Datasets and DataFrames - affect resource usage. Common applications of Spark analytics, such as counting, caching, partitioning, and KMeans, will be implemented using each of these datatypes and then will be run with different resource settings to see how the resource usage is affected. Since the current common benchmark of Spark analytics is execution time, this metric will also be analyzed to point out the downsides of being used as a benchmark for big data analytics.

Databricks, a company that supports the development of Spark and Spark related applications, is aware of the importance of resource usage. In an article analyzing the pros and cons of the different underlying datatypes of Spark analytics, they do bring up the fact that when a DataFrame is cached, it takes up significantly less memory than an RDD. [1] However, this is as far as they go in their analysis of Spark jobs from the perspective of resource usage.

This paper will take the analysis of resource usage by Spark jobs further by implementing multiple analytics and examining the peak execution memory metric, discussed further in the next section. Each of these jobs will be written as simply as possible to mimic the work a new Spark analytic developer would produce.

### A. SparkMeasure and Spark 2.4.0

The code written to accompany this paper was written for Spark 2.1.0, which is an older version of Spark. A library, written by an engineer at CERN named Luca Canali, will grab performance and resource metrics from the tasks run in a Spark job. This library is called SparkMeasure. [8] Since many Spark jobs have thousands, if not hundreds of thousands, of tasks SparkMeasure will also aggregate these metrics per Stage in order to create reasonably sized output files. In the jobs below, task metrics were captured for the jobs run on a local machine, while stage metrics were captured for the cluster run jobs.

The metric this paper will be focusing on is the peak execution memory since that is the most indicative of the resource usage of the Spark job.

Note that these metrics are not easily accessible in Spark 2.1.0; however, they were introduced in Spark 2.4.0 which was released November 2, 2018. [9]

## II. SPARK DISCUSSION RDD VS DATASET VS DATAFRAME

There are several differences between the three data formats that were used in this project. Resilient Distributed Dataset (RDD) data is an immutable representation of distributed data. RDDs are best for unstructured data such as text and byte streams. Datasets in the simplest terms are extensions of the Spark dataframe and provide an object oriented interface. DataFrames organize data into columns and provide a domain-specific API to manipulate distributed data. Data can be queried from both DataFrames and RDD through Spark SQL. [1]

Regarding the reported performance of the three data formats from other sources, RDDs offer a data format solution that is less expensive than the other data formats. This is, however, at the sacrifice of data performance. For a more efficient data format, DataFrames and Datasets provide a means of attaining greater optimization. [3] As specified in later sections, mapping and map partitions are optimizations that can be manually added to RDD for efficiency. Datasets in particular provide optimized queries through the Catalyst Query Optimizer (execution agnostic framework). The above performance notes will be apparent in the visualizations and analysis of the graphs in the pages that follow.

III. DATA DISCUSSION

The data set used is the Global Surface Summary of the Day (GSOD) which contains data, such as mean temperature, dew point, sea level pressure, mean wind speed, etc. Each data file thoroughly covers approximately a 24-hours period. [7] A cleanup was necessary because the data was not in an easily readable format, such as CSV or JSON. Having the raw data and transforming this data to CSV format ensured fair results for each of the 3 main data formats used in the project, and it enabled us to use built-in CSV parsing functionality. Further data processing involves removing the headers for the data since the headers were mostly incomplete. The dataset in total is 2.6GB large with 121,467,184 records. The mere size of the data posed a challenge on some of the machines that were used in the initial testing.

IV. SETUP

The jobs that were run on a local machine are from a MacBook Pro with a 2.2 GHz Intel Core i7 processor and 16 GB of 1600 MHz DDR3 memory. The cluster provided was fairly small and was in use by other individuals, thus, only about 40 GB and 8 cores were available for the use of this project.

V. MAP VS MAPPARTITIONS IN RDD

While writing Spark jobs, it is important to implement code that is efficient. The Spark jobs written for the purpose of this paper are not necessarily the most efficient, especially since the definition of efficient is somewhat subjective. This paper explores one definition of efficiency by examining the usage of available resources. Thus, the question of using the map method versus the mapParititions method is a good starting to place since it will define how the remaining Spark jobs using RDDs get written.

One of the most popular efficiencies is the usage of the mapParititions method instead of the map method when transforming an RDD. The map method will apply the specified transform to every element in the RDD. The mapPartitions method operates on a partition of the RDD. Simply put, a partition is a chunk of data in the RDD that can be operated on all at once without needing to shuffle the data. So, a mapPartitions method may apply the same transform to every data element within its partition, thus acting like the map method would. Keep in mind that mapPartitions may be used for more complicated operations utilizing all of the data elements within a partition.

While there are no traditional publications discussing the pros and cons of mapPartitions vs map, the general consensus on popular internet forums is that mapPartitions is generally faster because you do not have to set up a function for every element, just for the partition. [4]–[6] Since this section is not directly related to the purpose of this paper, a full written analysis will not be performed; however, this section provides information that fits into the larger purpose of this paper.

| Job Num | Threads | Executor Mem | Mem Overhead |
|---|---|---|---|
| 0 | 1 | 512m | 384m |
| 1 | 1 | 1024m | 384m |
| 2 | 1 | 1024m | 768m |
| 3 | 1 | 2048m | 768m |
| 4 | 2 | 512m | 384m |
| 5 | 2 | 1024m | 384m |
| 6 | 2 | 1024m | 768m |
| 7 | 2 | 2048m | 768m |
| 8 | 3 | 512m | 384m |
| 9 | 3 | 1024m | 384m |
| 10 | 3 | 1024m | 768m |
| 11 | 3 | 2048m | 768m |
| 12 | 4 | 512m | 384m |
| 13 | 4 | 1024m | 384m |
| 14 | 4 | 1024m | 768m |
| 15 | 4 | 2048m | 768m |

Fig. 1. List of Job Numbers and their respective resource settings for local machine runs.

A. Running on Local Machine

When running on a local machine, the resources that can be given to a Spark job are limited. Figure 1 provides information on the resource settings for local machine runs. This figure will be referenced in the other sections as well. In order to ensure that the job would not run out of resources, a subset of the NOAA GSOD dataset was selected. This subset of the NOAA GSOD data was everything from 2018. This dataset consisted of 3,443,046 records totaling ≈ 74 megabytes.

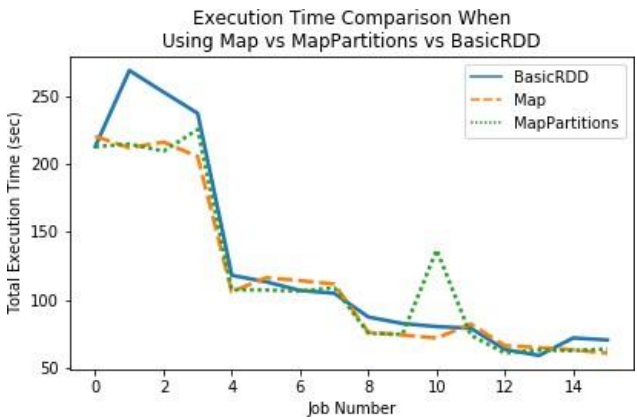

Fig. 2. Comparison of Map job vs MapPartitions job vs Basic RDD job

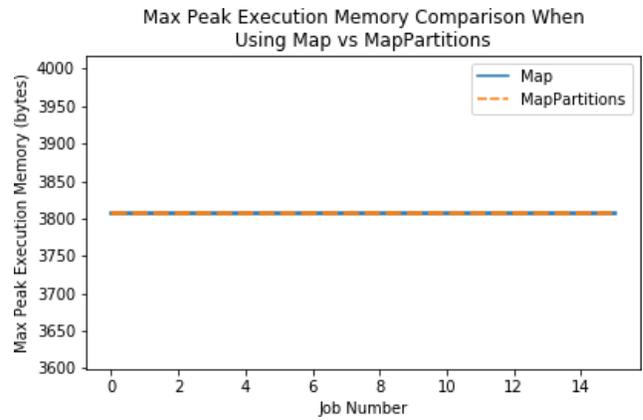

Fig. 3. Comparison of maximum peak execution memory of a Map job vs MapPartitions job runs on a local machine.

*1) Execution Time Comparison:* The execution time of the Spark job can be found by subtracting the minimum value of the launchTime stat from the maximum of the finishTime stat.

The focus of this paper is on the importance of using benchmarks other than execution time for Spark analytics and Figure 2 exemplifies this motivation by showing that:

- Even the exact same Spark analytic, such as the Basic RDD job and the MapPartitions job may not consistently have the same execution time.
- There can be anomalies in the execution time, which is likely due to the interference of other processes running on the same machine.

Additionally, Figure 2 also shows that there is, as expected, a relationship between the number of threads running the job and the execution time. For instance, doubling the threads from 1 to 2 cut the execution time in half. Tripling the threads from 1 to 3 reduced the execution time to about a third of the execution time of 1 thread. Increasing the number of threads from 1 to 4 also resulted in the run time decreasing to about 25% of the original run time.

This confirms that increasing the parallelization of the job by increasing the number of threads will decrease the run time; however, this does not confirm the commonly held belief that the mapPartitions method will execute faster than map method. This transitions to the next question analyzed of whether there is an advantage when it comes to the peak execution memory or not.

*2) Peak Execution Memory Comparison:* Figure 3 shows that the peak execution memory is the same for both map and mapPartitions.

### B. Running on Distributed Cluster

The cluster these Spark jobs were run on was not very large and other jobs were running at the same time. This is not a problem for our case since this paper is exploring the idea of resource benchmarking, and once Spark was given resources, they were not taken away. Figure 4 provides information on the resource settings for the cluster runs. This figure will be referenced in the other sections as well. These jobs ran over all of the NOAA GSOD dataset. This dataset consisted of 121,467,184 records totaling ≈ 2.6 gigabytes.

| Job Num | Num Executors | Num Cores | Executor Mem | Mem Overhead |
|---|---|---|---|---|
| 0 | 3 | 1 | 3072m | 2048m |
| 1 | 3 | 2 | 3072m | 2048m |
| 2 | 4 | 1 | 3072m | 2048m |
| 3 | 4 | 2 | 3072m | 2048m |

Fig. 4. List of Job Numbers and their respective resource settings for cluster runs.

Note that the jobs were not able to be run using only one or two executors due to cluster limitations on the maximum size of a container given to executors. Similarly, the cluster could not support a higher number of executors.

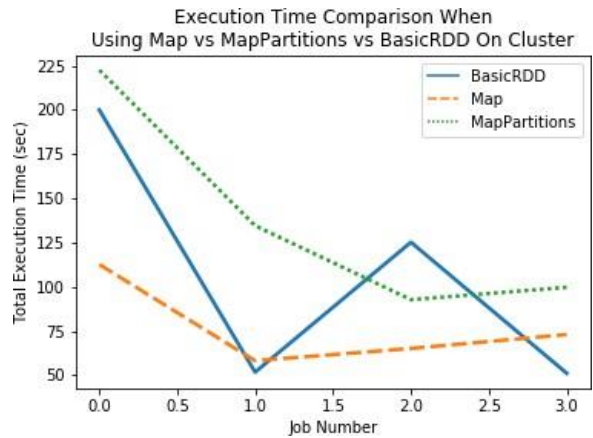

Fig. 5. Comparison of the execution time of a Map job vs MapPartitions job vs Basic RDD job runs on the cluster

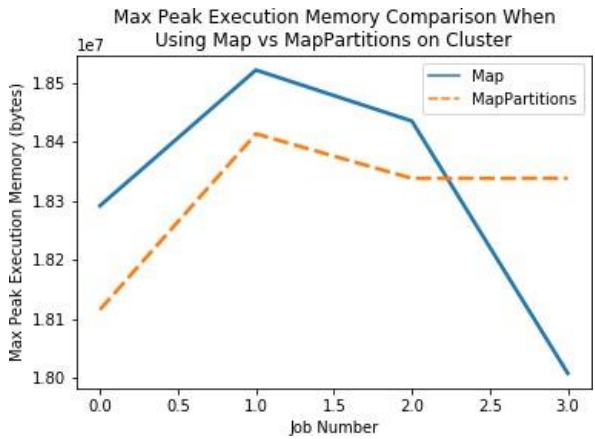

Fig. 6. Comparison of maximum peak execution memory of a Map job vs MapPartitions job runs on the cluster.

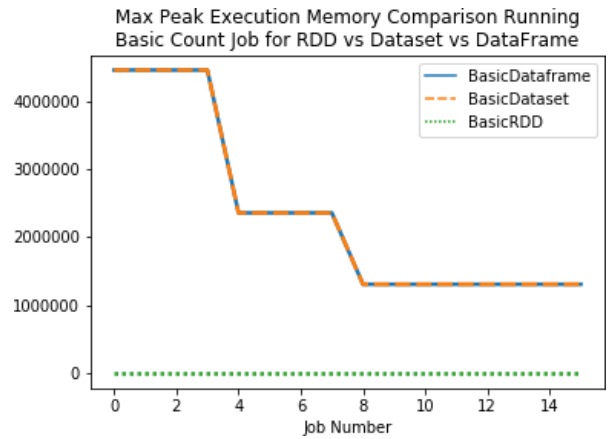

Fig. 7. Comparison of maximum peak execution memory of the basic job runs on a local machine.

*1) Execution Time Comparison:* Again, Figure 5 exhibits how benchmarking Spark jobs based only on execution time does not give the entire picture. The Basic RDD and MapPartitions job are the exact same and yet, Figure 5 shows the run times are not consistent.

*2) Peak Execution Memory Comparison:* By examining Figure 6, it can be seen that in 3/4 of the runs the mapPartitions method had a lower amount of peak execution memory. Also note that the peak execution memory metric for the cluster runs is significantly higher than that of the local machine runs. This makes sense because the size of the dataset was significantly larger.

Since it has been established above that the peak execution memory during a job using mapPartitions will be equal to the job using the map method on a local machine and will be lower for the majority of runs in the cluster, the remaining RDD jobs are written using mapPartitions.

## VI. BASIC RDD VS DATASET VS DATAFRAME

The equivalent of the Hello World program for any big data analytics platform, including Spark, is Word Count. The basic premise of the Word Count program is to count all of the unique words in a document. The NOAA GSOD data does not easily support a Word Count program, so instead, a similar program was created to start the analysis of the peak execution memory for RDDs, DataFrames and Datasets. In this job, a key was created by combining the station number and the nearest multiple of 10 of the average temperature that does not go over the average temperature (Ex 98 $\rightarrow$ 90). Each unique key is then counted. The output of the program is a count of all of the unique keys as this will force the program to run in its entirety to ensure equal comparison.

### A. Running on Local Machine

The description of the resource settings for the runs on the local machine is the same as above and can be seen in Figure 1.

*1) Peak Execution Memory Comparison:* As it can be seen in Figure 7, the maximum peak execution memory for the DataFrame and Dataset implementations are the same. This is expected since they are utilizing the same type of optimization behind the scenes. Additionally, Figure 7 shows a step like behavior for the peak execution memory for these two jobs based on the thread count. The first four jobs are run using one thread, and thus, have a higher peak execution memory. Upon examination of the job using the Spark UI, it can be seen that one of the optimizations of the Dataframes and Datasets is that it will repartition the data. This will be discussed in further detail in Section VII, but simply put, this means that more data is being put into a partition, so more memory is needed to deal with a single partition. So, the next four jobs are run using two threads, so more partitions are created compared to running the job with one thread. Thus, each partition is smaller in the two-threaded job than in the one threaded job. Similarly, this also explains the step caused in the next 8 jobs when three and four threads are used. The current working theory as to why there is not a step between the 3-threaded job and the 4-threaded job is because partitions are created based on the number of cores being used. The local job was run on a machine with a quad-core processor, so one of the processors may have been assigned to the driver.

This RDD implementation also has a significantly lower peak execution memory, sitting at $\approx$ 4000 bytes. The peak execution memory is so low, comparatively, for the RDD implementation because it does not have any optimization going on behind the scenes. For each partition of data that the job has read in, it processes that partition individually during the count stage. Thus, the peak execution memory remains the same.

This leads to the question of how does forcing the RDD to repartition affect the peak execution memory. This will be explored in the next section.

*2) Execution Time Comparison:* As we can see above, RDDs will use a lower amount of execution memory.

However, the graph in Figure 8 shows that RDDs are also slower in the case of these jobs. This makes sense because the overhead of preparing an executor to work with a partition is not insignificant. Since the number of partitions that an RDD needs to work with is higher, it will take longer due this overhead. Note that there were about 12000 partitions in the RDD during the count stage, while the DataFrame and Datasets were only working with at most 500 partitions.

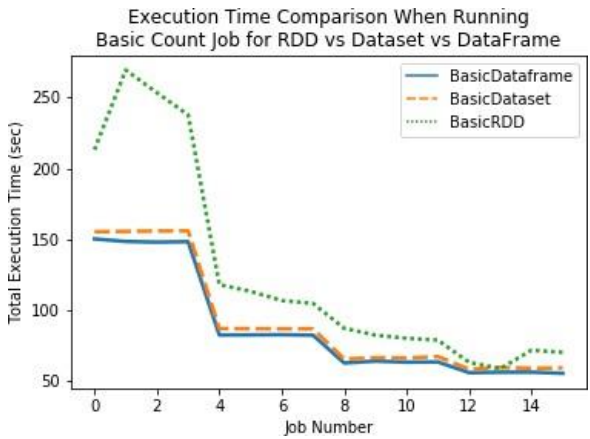

Fig. 8. Comparison of execution time of the basic job runs on a local machine.

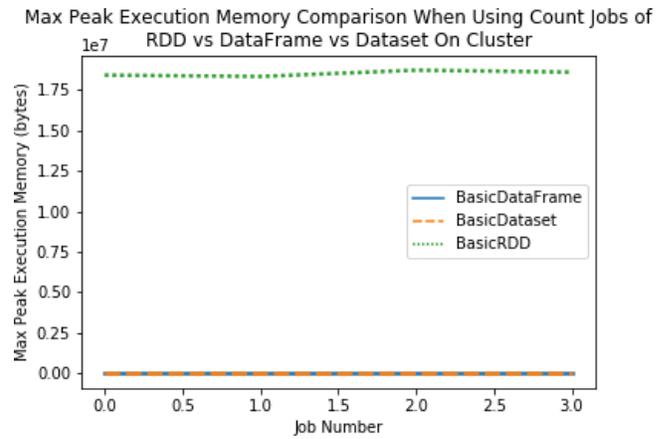

Fig. 9. Comparison of maximum peak execution memory of the basic job runs on a cluster.

Interestingly enough, the gap between the run times of the RDD vs DataFrames/Datasets closed as more threads were used. This may indicate that at some point, having enough cores to run the job against removes the speed benefit of DataFrames and Datasets; however, exploring this is out of scope.

### B. Running on Distributed Cluster

Unfortunately, due to the clusters limited resources, the DataFrame and Dataset implementations of this basic count were unable to successfully complete with the configuration provided. This does imply that their peak execution memory would be too high since the message displayed as to why the DataFrame and Datasets jobs died were because they used all of the resources and needed more.

Note that these jobs used the same resources settings as shown in Figure 4.

*1) Peak Execution Memory Comparison:* Figure 9 shows that again, the maximum peak execution memory used is constant regardless of the number of executors or core per executor. This makes sense because for this job with RDDs, the same partitions are going to be processed the same way, so the peak execution memory shouldn't flux much, if at all.

*2) Execution Time Comparison:* Figure 10 shows that the in both the 3 and 4 executor cases, having double the cores per executor significantly speeds up the execution time.

### VII. PARTITIONS RDD VS DATASET VS DATAFRAME

The partition jobs are basically copies of the "Basic" jobs discussed in Section VI; however, the coalesce method was used on the raw input. The coalesce method was used instead of the repartition method because repartition will force a shuffle. The coalesce method did not force a shuffle and will instead just redistribute the old partitions to the specified number of new partitions. Coalesce was the chosen method because even in the smaller job running on a local machine, over 12000 partitions were going to be created, so reducing the number of partitions should not require a shuffle to redistribute the data.

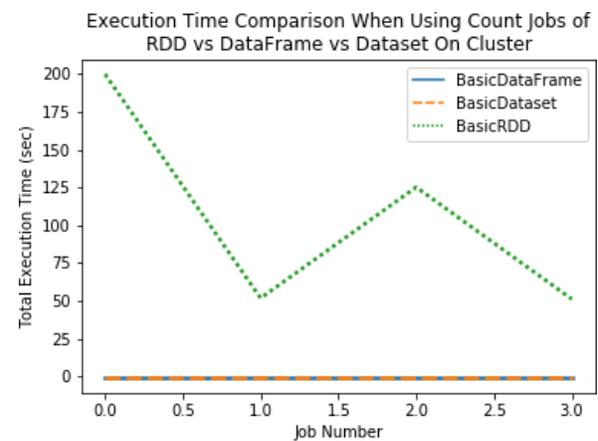

Fig. 10. Comparison of execution time of the basic job runs on a cluster.

### A. Running on Local Machine

Figure 11 shows the resource configurations of the cluster jobs for the partition jobs being run on a local machine.

*1) Resource Comparison for different partitions:* Like the previous job examples, we can see that again in all cases the RDD will have a lower peak execution memory, even when the partitions are the same. Figure 12 makes some of the PartitionRDD jobs look like they did not run and have a peak execution memory of 0. The RDD jobs did run and the max peak execution memory for the PartitionRDD has a min of 16000 bytes.

Figure 12 also shows that for every underlying datatype, the number of partitions has a large effect of the peak execution memory. For all of them, the highest peak execution memory occurred when the data was coalesced to one partition, which makes sense because then all of the data is being worked with at once. Interestingly, the difference in peak execution memory when using 100 vs 500 vs 1000 partitions was almost negligible. This would indicate that there is not much benefit in using many more than 100 partitions for these jobs in terms of trying to lower the peak execution memory.

| Job Num | Threads | Executor Mem | Mem Overhead | Partitions |
|---|---|---|---|---|
| 0 | 1 | 512m | 384m | 1 |
| 1 | 1 | 1024m | 384m | 100 |
| 2 | 1 | 1024m | 768m | 500 |
| 3 | 1 | 2048m | 768m | 1000 |
| 4 | 2 | 512m | 384m | 1 |
| 5 | 2 | 512m | 384m | 2 |
| 6 | 2 | 1024m | 384m | 100 |
| 7 | 2 | 1024m | 768m | 500 |
| 8 | 2 | 2048m | 768m | 1000 |
| 9 | 3 | 512m | 384m | 1 |
| 10 | 3 | 512m | 384m | 3 |
| 11 | 3 | 1024m | 384m | 100 |
| 12 | 3 | 1024m | 768m | 500 |
| 13 | 3 | 2048m | 768m | 1000 |
| 14 | 4 | 512m | 384m | 1 |
| 15 | 4 | 512m | 384m | 4 |
| 16 | 4 | 1024m | 384m | 100 |
| 17 | 4 | 1024m | 768m | 500 |
| 18 | 4 | 2048m | 768m | 1000 |

Fig. 11. List of Job Numbers and their respective resource settings for local runs of partition jobs.

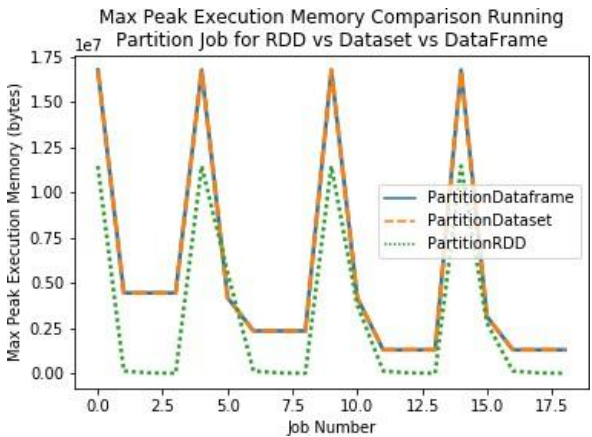

Fig. 12. Comparison of maximum peak execution memory of the partition job runs on a local machine.

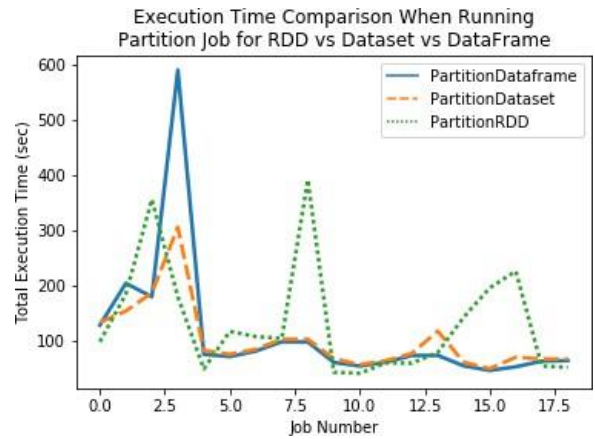

Fig. 13. Comparison of execution time of the partition job runs on a local machine.

| Job Num | Num Executors | Cores | Executor Mem | Mem Overhead | Partitions |
|---|---|---|---|---|---|
| 0 | 4 | 1 | 3072m | 2048m | 4 |
| 1 | 4 | 1 | 3072m | 2048m | 100 |
| 2 | 4 | 1 | 3072m | 2048m | 400 |
| 3 | 4 | 1 | 3072m | 2048m | 1000 |
| 4 | 4 | 2 | 3072m | 2048m | 4 |
| 5 | 4 | 2 | 3072m | 2048m | 100 |
| 6 | 4 | 2 | 3072m | 2048m | 400 |
| 7 | 4 | 2 | 3072m | 2048m | 1000 |

Fig. 14. List of Job Numbers and their respective resource settings for runs on a cluster.

*2) Execution Time Comparison:* Like the previous execution time graphs, Figure 13 shows why execution time is a bad benchmark. For example, most of the Dataset and DataFrame backed jobs have very similar run times; however, there is the random spike in run time at job 3.

*B. Running on Distributed Cluster*

Unfortunately, the PartitionRDD job was the only job that completed when running on the cluster. The PartitionDataset and PartitionDataFrame jobs were not able to complete because they ran out of resources.

Figure 11 shows the resource configurations of the cluster jobs for the partition jobs being run on a local machine.

*1) Resource Comparison for different partitions:* As it can be seen in Figure 15, job numbers 3 and 7 of the Partition-

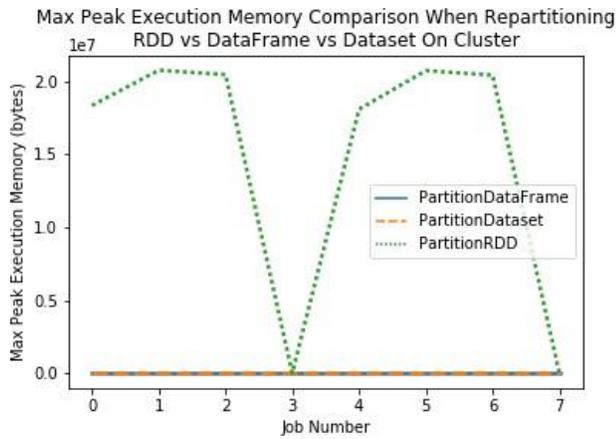

Fig. 15. Comparison of peak execution memory of the partition job runs on a cluster.

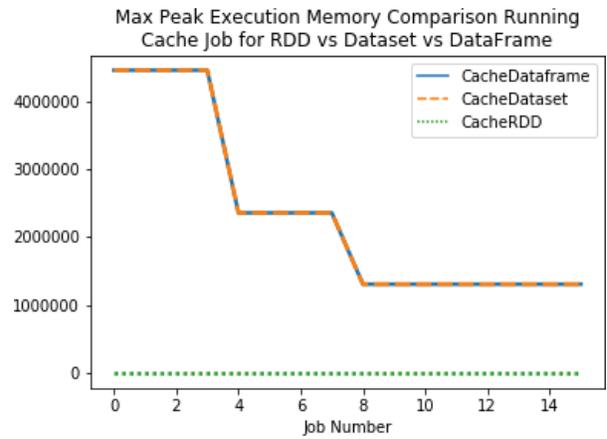

Fig. 16. Comparison of maximum peak execution memory of the caching job runs on a local machine.

RDD job also failed due to lack of resources. An interesting thing that can be seen in this graph is that the maximum peak execution memory actually increases as the number of partitions increase. This is surprising since the behavior was the opposite on the local run. In the local runs of this job, the peak execution memory decreased as the number of partitions increased. This is possibly because of the coalesce behavior. This job was trying to get the ≈400000 partitions read in condensed into a much smaller number of partitions. We can also compare this peak execution memory to that in Figure 9 for the MapPartitions job because that job just worked with the original partitions instead of trying to coalesce. The peak execution memory of the MapPartitions job is smaller than that of the PartitionRDD job, which is likely due to the fact that the MapPartitions job is working with smaller clusters.

Additionally, it is important to note that the difference in the peak execution memory for the jobs running with only one core per executor and those running with two cores per executors is almost negligible.

*2) Execution Time Comparison:* Since the PartitionDataset and PartitionDataFrame jobs did not complete, this section will not be analyzed since no comparison can be made.

## VIII. CACHE RDD VS DATASET VS DATAFRAME

The caching jobs are similar to the basic jobs discussed in Section VI; however, the data is cached after it is read, but before there is any manipulation of the data. This reflects common use cases of caching the raw data before manipulating it multiple ways.

There were issues running all of the caching jobs on the cluster since every single job ran out of resources. Thus, only the case where the jobs were run on a local machine will be analyzed.

*1) Resource Comparison for different partitions:* Yet again, Figure 16 exhibits the fact that RDDs have a lower peak execution memory than DataFrames or Datasets. This graph does match that in Figure 7, which makes sense because they are basically the same job. This does imply that the peak execution memory is heavily affected by the partition size and not so much the data being stored in memory.

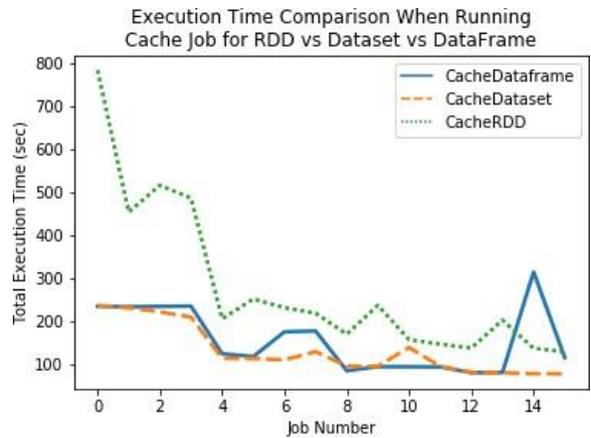

Fig. 17. Comparison of execution time of the partition job runs on a local machine.

*2) Execution Time Comparison:* Figure 17 again exemplifies how the execution time is not guaranteed to be a reproducible metric, which means it is not an ideal benchmark.

## IX. KMEANS RDD VS DATASET VS DATAFRAME

KMeans is a popular machine learning algorithm and Apache Spark has a built-in implementation of it for RDDs, Datasets and DataFrames. Surprisingly, this single job caused the RDD implementation to exceed its assigned resources while the Dataset and DataFrame versions were able to finish. This is likely due to the fact that the RDD implementation is from Spark version 1.* whereas the Dataset and DataFrame implementations are being updated. Additionally, all of the KMeans jobs failed to run on the cluster due to running out of resources.

The run settings for the KMeans jobs are the same as those in Figure 1.

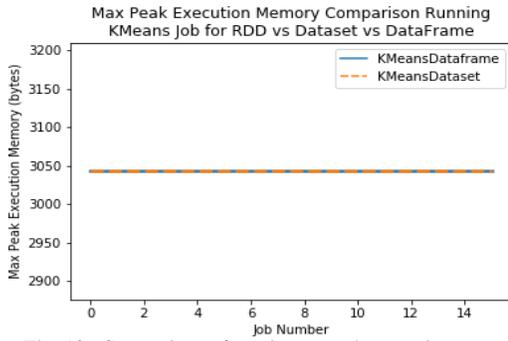

Fig. 18. Comparison of maximum peak execution memory of the kmeans job runs on a local machine.

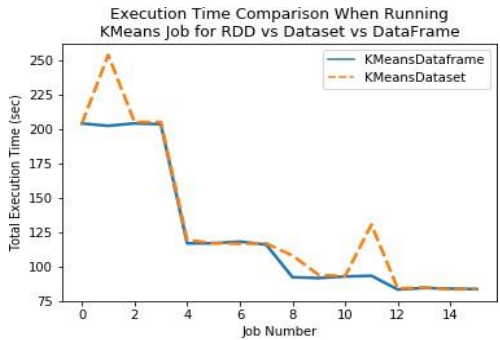

Fig. 19. Comparison of execution time of the kmeans job runs on a local machine.

1) *Resource Comparison for different partitions:* The maximum peak execution memory for the Dataset and DataFrame Kmeans jobs is constant. This behavior is similar to that shown in Figure 7, which is not surprising since KMeans is basically going to iterate over all of the partitions while calculating the cluster centers, which is what the BasicRDD job does.
2) *Execution Time Comparison:* The execution time for the KMeans jobs are provided in Figure 19. There is nothing out- standing in this figure beyond seeing that there are unexplained anomalies in the execution time for the KMeansDataset job.

## X. RELATED WORK

Several research works have used different forms of benchmark to evaluate the performance of the systems design.

Researchers with a focus of resource management in the area of sensor network [10-24] use the metrics such as error rate, and energy consumption to evaluate the scheduling quality of a sensor network. The other metrics include delivery ratio, and normalized number of transmissions which reflects the energy efficient of a forwarding protocol. The results from the sensor network research show that simulation-based techniques are considered as effective methods to evaluate the performance of network protocols.

Researchers that focus on resource usage and management in smart grids [25-30] use other metrics, such as Mean Absolute Percentage Error (MAPE) of peak demand and MAPE of hourly average power to analyze the prediction the energy consumption in smart grids.

In the data management and forwarding [31-35], there are various of methods such as Amphista to facilitate concurrent uplink and downlink efficient communication among the Internet of Things (IoT) devices along with other huge amount of data.

Researchers in security related problems [36-40] have considered other form of privacy protection framework such as Shepherd that can utilize the renewable energy while still preserving privacy since it effectively hides the power consumption information.

Recent studies based on energy and power consumption data [41-50] have utilized metrics that includes storage space and average energy or power consumption in order to evaluate the effectiveness of a middleware E-Sketch and Passive Zigbee or backscatter to relay data between WiFi to Zigbee devices.

The main contribution of this paper is to resolve the problem of relying on execution time metric to evaluate resource usage. The peak memory execution metric was established to be an ideal metric to benchmark the resource usage while analyzing the Spark jobs.

## XI. CONCLUSION

Through the exploration of the peak execution memory and execution time metrics of the 5 types of jobs above, this paper has determined the following:
- RDDs consistently have a lower peak execution memory than DataFrames and Datasets. Thus, they are more stable and frequently jobs implemented using RDDs would complete when the same job would not complete if it was implemented using DataFrames or Datasets.
- The peak execution memory is most heavily affected by the size of partitions.
- Jobs with more data have a higher peak execution memory than the exact same job run over less data.
- Jobs with fewer partitions have a higher peak execution memory
- The peak execution memory does not seem to be affected by the numbers of threads, executors, or cores.
- The execution time is not guaranteed to be reproducible. The same job run twice can have different run times.
- The execution time is most affected by the number of cores, executors, and threads.
- Adding more memory resources to a job that can finish with less will not change the peak execution memory. This action also does not significantly reduce run time.

Ultimately, while Databricks [1] and Apache Spark [9] push DataFrames and Datasets as being optimized, this optimization tended to need more resources. These optimized jobs tended to finish more quickly than jobs using RDDs; however, RDDs used less resources at a time, and thus, would successfully complete the job much more often.

Additionally, all this analysis supports the point that peak execution memory of the resource usage is an important benchmark to include when comparing frameworks for big data analytics. While execution time is important, clearly conveying how a job is affected by the resources available is a critical component when comparing analytic jobs.